\documentclass[epsfig]{article}

\usepackage{graphicx}

\begin{document}

%\markboth{J.R. Pel\'aez}
%{Regge description of high energy pion pion total cross sections}

%%%%%%%%%%%%%%%%%%%%% Publisher's Area please ignore %%%%%%%%%%%%%%%
%
%\catchline{Regge description of $\pi\pi$ total cross sections}{}{}{}{}
%
%%%%%%%%%%%%%%%%%%%%%%%%%%%%%%%%%%%%%%%%%%%%%%%%%%%%%%%%%%%%%%%%%%%%

\title{Regge description of high energy pion pion total cross sections
\footnotetext{To appear in the proceedings of MESON2004, Krakow, Poland, July 2004.}}

\author{ Jos\'e R. Pel\'aez
\\Departamento de F\'{\i}sica Te\'orica II.
Universidad Complutense. 28040 Madrid. Spain.
}

\maketitle

\begin{abstract}
We have recently presented a Regge description of
$\pi\pi$ total cross sections valid above 1.4 GeV,
consistent with the few existing experiments, 
factorization and crossing symmetry.
In this note we show how it also describes a further large data sample
obtained from an analysis of experiments on $\pi^\pm p\rightarrow X\Delta^{++}$
and $\pi^\pm n\rightarrow Xp$.
%\keywords{Regge Theory; pion scattering.}
\end{abstract}

\section{Regge description of $\pi\pi$ total cross sections}    %) A SECTION HEADING

In references  \cite{nos1,nos2}, we have shown how it was possible
to obtain a precise Regge description of high energy total $\pi\pi$ scattering
down to $E_{kin}\simeq1.1$ GeV. 
Apart from the interest in itself,
there has been a renewed interest in this high energy region because
the imaginary part of the $\pi\pi\rightarrow\pi\pi$ amplitude 
is needed for dispersive studies aiming at a precise
description of $\pi\pi$ data at low energies  \cite{ACGL,disp,nos1}.

A relevant property of our
description is that it respects {\it factorization}.
This means that, for instance,
the imaginary part of an amplitude $F_{A+B\rightarrow A+B}$ is:
\begin{equation}
  \label{eq:factorization}
  \hbox{Im}\, F_{A+B\rightarrow A+B}(s,t)\simeq f_A(t) f_B(t) (s/\hat s )^{\alpha_R(t)},
\quad \hat s= (1\, \hbox{GeV})^2.
\end{equation}
The $ (s/\hat s )^{\alpha_R(t)}$ behavior comes from the so-called Regge pole $R$.
All poles have $\alpha_R<1$ and thus vanish for large $s$,
except the Pomeron that scales like $s$ up
to around 15 or 20 MeV, where it dominates all other
pole contributions, giving a common prediction $\sigma^\infty$
for all $\pi\pi$ channels. For larger energies it increases logarithmically.
As a matter of fact there could
be many Regge poles exchanged in each channel, all them with their corresponding
$f_i(t)$ factors depending on $R$  and the particles in the initial state.
Using factorization, it is thus possible to obtain the $\pi\pi$ Regge amplitudes 
from those of $\pi N$ and $NN$. Total cross sections
are then related to forward scattering amplitudes by:
$\sigma_{AB}=4\pi^ 2 \hbox{Im}\, F_{A+B\rightarrow A+B}(s,0)
/\lambda^{1/2}(s,m_A^2,m_B^2)$, with $\lambda(a,b,c)=a^2+b^2+c^2-2ab-2ac-2bc$.
Thus we \cite{nos2}  fitted the large
$\pi^\pm N$ and $NN$ data compilation of the COMPASS
group as given in the Particle Data Tables \cite{PDG}, and the few $\pi\pi$
data \cite{totaldata} points known to us down to $E_{kin}\simeq1.1$ 
imposing factorization. The fit parameters are largely dominated
by the $\pi^\pm N$ and $NN$ experiments, but
still we obtained a very precise description for $\pi\pi$
total cross sections, that was in remarkable agreement with
the $\sigma_{tot}^{\pi\pi}$ data above 2 GeV. At lower energies these
data are in conflict with the $\sigma_{tot}^{\pi\pi}$ reconstructed \cite{nos2} from
lower energy phase shifts analysis and our
results fall somewhere in-between. 
We refer to our 
paper \cite{nos2} for further details.

In addition, we have also checked that our high energy 
results 
together with fits \cite{nos1} to the low energy
satisfy two crossing symmetry sum rules. 
This is again of relevance
because in the seventies \cite{Pennington1} there was a suggestion that
the predictions of factorization $\sigma^\infty\simeq 13\,$mb, 
together with the
existing phase shifts analysis at that time, violated crossing symmetry,
suggesting $\sigma^\infty= 6\pm5\,$mb.
Of course this was tenable until the first high energy
$\sigma_{tot}^{\pi\pi}$ were measured, 
and indeed the very same authors \cite{Pennington1} pointed out somewhat later
that the central
value should be raised to $\sigma^\infty=8.3\,$mb.
The recent studies in  \cite{ACGL,disp} 
used $\sigma^\infty= 5\pm3\,$mb, following
 \cite{ACGL}.
Unfortunately, the $\sigma_{tot}^{\pi\pi}$ data went largely unnoticed 
to our days, including to ourselves, so that in  \cite{nos1} 
the use of factorization was only based
on QCD considerations.  In  \cite{nos2} we ``rediscovered'' 
four different experimental works \cite{totaldata} that we used in a reanalysis
to find 
$\sigma_{tot}(20\hbox{GeV})=13.4\pm 0.6\,$mb, while simultaneously
respecting crossing.

\section{Comparison with further data}

Following the discussions of my talk on this MESON2004 conference
I came to know that there was another analysis \cite{Zakharov} of $\pi \pi$ 
total cross sections. In that work, a triple reggeon model is used to analyze
several sets of experimental data on $p p\rightarrow X \Delta^{++}$
and $p n (p)\rightarrow X p (n)$, and obtain Regge parameters with whom 
to extract total $\pi^\pm\pi^-$
cross sections from $\pi^\pm p\rightarrow X \Delta^{++}$ and
$\pi^\pm n\rightarrow Xp$. 
The most relevant contribution of this paper is the inclusion
of absorptive corrections in the last two reactions, which seems to decrease
the results by about 10 to 15\%. In Fig.1, we show how our
Regge description, and in particular, our value 
$\sigma_{tot}(20\hbox{GeV})=13.4\pm 0.6\,$mb
 indeed provides a good description of this data, which strongly disfavors
a value more than two times smaller. Following the authors we display only
the statistical errors. Systematic errors were estimated at the $7-10\%$
level.

\begin{figure}
%\centerline{\psfig{file=fig1pelaez.eps,width=9.cm}}
%\centerline{\psfig{file=fig2pelaez.eps,width=9.cm}}
\centerline{\includegraphics[scale=.75]{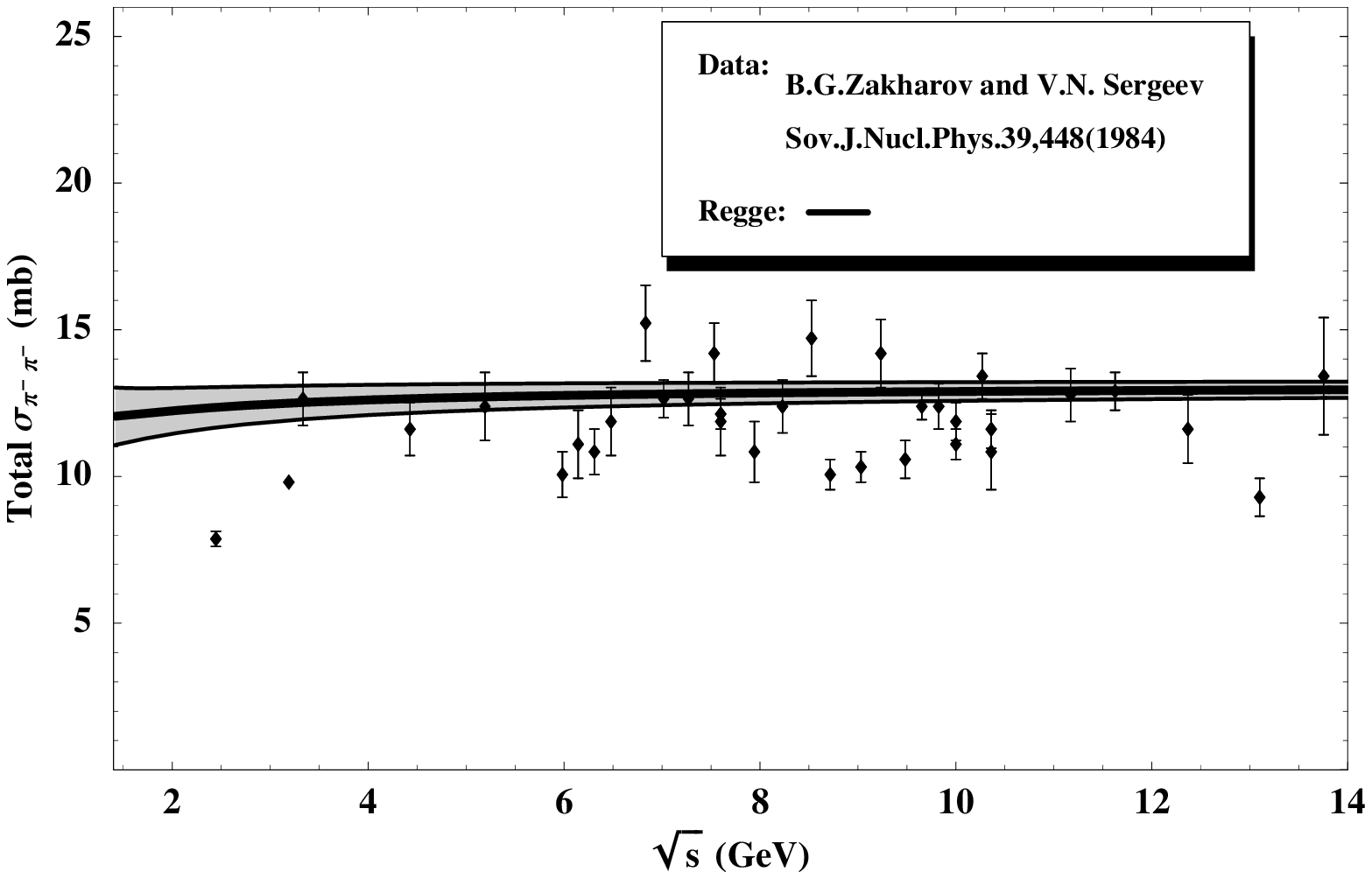}}
\centerline{\includegraphics[scale=.75]{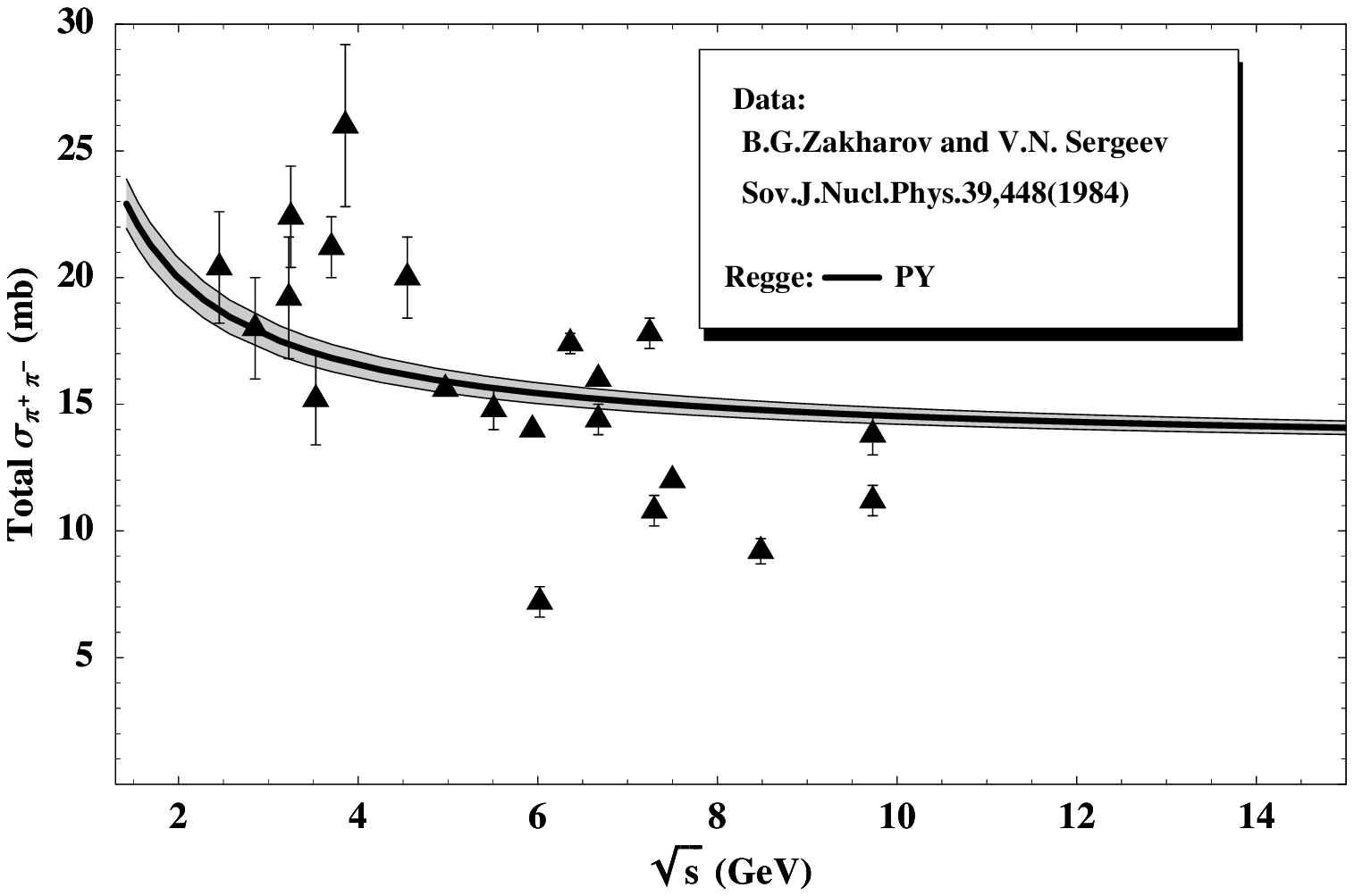}}
\vspace*{8pt}
\caption{The continuous line stands for our Regge representation
and the gray band for the associated uncertainty. Data are from [8]
and the error bars are just statistical, however, the authors
pointed out a ``possible systematic error of $\simeq7-10\%$``.}
\end{figure}

\section*{Acknowledgments}

I thank the MESON2004 organizers for the stimulating 
workshop, since this note originated in the
discussion following my talk. In particular, I thank 
A. Szczurek and N.N. Nikolaev for their comments
on the data \cite{Zakharov}.
I also thank F.J. Yndur\'ain for comments and suggestions
and as coauthor of the theoretical
Regge description reviewed here.
I am very grateful for the hospitality of the 
Institut f\"ur Kernphysik (Theorie), Forschungzentrum J\"ulich,
where this note was prepared. Financial support from
Spanish CICYT projects BFM2000-1326, BFM2002-01003,
and from the E.U. EURIDICE network HPRN-CT-2002-00311
is also acknowledged.


\begin{thebibliography}{0}

\bibitem{nos1} 
J.~R.~Pel\'aez and F.~J.~Yndur\'ain,
%``On the precision of chiral-dispersive calculations of pi pi scattering,''
\textsl{Phys.\ Rev.} {\bf D68}, 074005 (2003)
%[arXiv:hep-ph/0304067].
%%CITATION = HEP-PH 0304067;%%

\bibitem{nos2} 
% \cite{Pelaez:2003ky}
%\bibitem{Pelaez:2003ky}
J.~R.~Pel\'aez and F.~J.~Yndur\'ain,
%``Regge analysis of pion pion (and pion kaon) scattering for energy s(1/2) >
%1.4-GeV,''
\textit{Phys.\ Rev.\ } {\bf D69}, 114001 (2004)
%[arXiv:hep-ph/0312187].
%%CITATION = HEP-PH 0312187;%%

\bibitem{ACGL} B.Ananthanarayan, G.Colangelo, J.Gasser and H.Leutwyler,
%``Roy equation analysis of pi pi scattering,''
\textsl{Phys.\ Rept.\ } {\bf 353}, 207 (2001)
%[arXiv:hep-ph/0005297].
%%CITATION = HEP-PH 0005297;%%

\bibitem{disp}G.~Colangelo, J.~Gasser and H.~Leutwyler,
%``pi pi scattering,''
\textsl{Nucl.\ Phys.\ }B {\bf 603}, 125 (2001);
%[arXiv:hep-ph/0103088].
%%CITATION = HEP-PH 0103088;%%
% \cite{Descotes-Genon:2001tn}
%\bibitem{Descotes-Genon:2001tn}
S.~Descotes-Genon, N.~H.~Fuchs, L.~Girlanda and J.~Stern,
%``Analysis and interpretation of new low-energy pi pi scattering data,''
\textsl{Eur.\ Phys.\ J.\ }C {\bf 24}, 469 (2002);
%[arXiv:hep-ph/0112088].
%%CITATION = HEP-PH 0112088;%%
% \cite{Kaminski:2002pe}
%\bibitem{Kaminski:2002pe}
R.~Kaminski, L.~Lesniak and B.~Loiseau,
%``Elimination of ambiguities in pi pi phase shifts using crossing symmetry,''
\textsl{Phys.\ Lett.\  }B {\bf 551}, 241 (2003);
%[arXiv:hep-ph/0210334].
%%CITATION = HEP-PH 0210334;%%
% \cite{Buettiker:2003pp}
%\bibitem{Buettiker:2003pp}
P.~Buettiker, S.~Descotes-Genon and B.~Moussallam,
%``A re-analysis of pi K scattering a la Roy and Steiner,''
\textsl{Eur.\ Phys.\ J.}\ C {\bf 33}, 409 (2004).
%[arXiv:hep-ph/0310283].
%%CITATION = HEP-PH 0310283;%%


\bibitem{PDG} K. Hagiwara \textit{et al.}, \textit{Phys. Rev.} {\bf D66} 010001 (2002).

\bibitem{totaldata}Biswas, N. N., et al., {\sl Phys. Rev. Letters}, 
{\bf 18}, 273 (1967) [$\pi^-\pi^-$, $\pi^+\pi^-$ and $\pi^0\pi^-$];
 Cohen, D. et al., {\sl Phys. Rev.}
{\bf D7}, 661  (1973) [$\pi^-\pi^-$];
 Robertson, W. J.,
Walker, W. D., and Davis, J. L., {\sl Phys. Rev.} {\bf D7}, 2554  (1973)  [$\pi^+\pi^-$]; 
Hoogland, W., et al.  {\sl Nucl. Phys.}, {\bf B126}, 109 (1977) [$\pi^-\pi^-$];
Hanlon, J., et al,  {\sl Phys. Rev. Letters}, 
{\bf 37}, 967 (1976) [$\pi^+\pi^-$]; Abramowicz, H., et al. {\sl Nucl. Phys.}, 
{\bf B166}, 62 (1980) [$\pi^+\pi^-$].

\bibitem{Pennington1}M.~R.~Pennington,
 %``How Crossing Affects The Analytic Relationship Between Low-Energy Pi Pi
%Amplitudes And Their Asymptotic Behavior,''
\textsl{Annals Phys.}\  {\bf 92}, 164 (1975).
%%CITATION = APNYA,92,164;%%
A.~D.~Martin and M.~R.~Pennington,
 %``How Imposing Analyticity On A Pi Pi Phase Shift Analysis Can Reveal New
%Solutions, Explore Experimental Structures And Investigate The Possibility Of
%New Resonances,''
 \textsl{Annals Phys.}\  {\bf 114}, 1 (1978).
%%CITATION = APNYA,114,1;%%

\bibitem{Zakharov} B.G. Zakharov and V.N. Sergeev, {\it Sov. J. Nucl. Phys.} 
{\bf 39}, 448 (1984) also in Yad.\ Fiz.\  {\bf 39}, 707 (1984).
%%CITATION = YAFIA,39,707;%%
\end{thebibliography}
\end{document}